\documentstyle[osa,epsfig,multicol]{revtex}

\begin{document}

\title{\LARGE Dynamics of modal power distribution in a multimode
semiconductor laser with optical feedback}

\author{\bf J.M. Buld\'u,
J. Trull,
M.C. Torrent,
J. Garc\'{\i}a-Ojalvo
}
\address{Departament de F\'{\i}sica i Enginyeria Nuclear, Universitat
Polit\`ecnica de Catalunya, Colom 11, E-08222 Terrassa, Spain
}
\author{\bf
Claudio R. Mirasso
}
\address{
Departament de F\'{\i}sica, Universitat de les Illes Balears,
E-07071 Palma de Mallorca, Spain}

\maketitle

\begin{abstract}
The dynamics of power distribution between longitudinal modes of a
multimode semiconductor laser subjected to external optical feedback
is experimentally analyzed in the low-frequency fluctuation regime.
Power dropouts in the total light intensity
are invariably accompanied by sudden activations of
several longitudinal modes.
These activations are seen not to be simultaneous to the dropouts, but
to occur after them. The phenomenon is statistically analysed in a
systematic way, and the corresponding delay is estimated.

\end{abstract}


\begin{multicols}{2}
{

Semiconductor lasers are devices very susceptible to exhibiting unstable
dynamical behavior. When subjected to reflections of its own emitted
radiation, in particular, they easily enter
complex dynamical regimes, exhibiting for instance low-frequency
fluctuations (LFF) in the form of
intensity dropouts \cite{risch}, or fully-developed chaotic fluctuations
leading to coherence collapse \cite{lenstra}. Most of the
related theoretical and experimental studies undertaken so far have
dealt with the dynamics of the total emitted intensity
\cite{lff,claudio}.
However, the low-cost semiconductor
lasers employed in technological applications operate usually in several
longitudinal modes. Therefore, analysing the mode dynamics would be
necessary, a need which has
been recognized only recently \cite{ryan}. In particular, recent
experiments have indeed shown the importance of multimode operation
in the LFF regime \cite{huyet,vaschenko}.
Different dynamical \cite{mandel,rogister} and statistical
\cite{sukow} characteristics of this regime
have been described in terms of a multimode extension of the well-known
Lang-Kobayashi model \cite{lang}.

In the course of the above-mentioned investigations, it was observed that
when the feedback was frequency-selective (such as that provided by a
diffraction grating), the intensity dropouts were accompanied by
a sudden activation of other longitudinal modes of the laser
\cite{huyet,giudici}. These modes,
located at the sides of the main mode in the gain curve, will be
called longitudinal side modes, or simply side modes, in the rest of
this Letter.
The activation of these modes was heuristically interpreted
as the mechanism producing the intensity dropouts \cite{giudici}, and
was numerically reproduced again by a multimode LK model \cite{mandel2}.
In this Letter, we show experimentally that the side-mode
activation also appears in the presence of non-frequency-selective
feedback, and that it occurs neither simultaneously nor previously
to the intensity dropout, but {\em after} it. Therefore, in this
case this activation can not be the cause, but rather the effect,
of the dropout.

Our experimental setup is shown schematically in Fig. \ref{fig:setup}.
We use an index-guided single-transverse-mode
AlGaInP semiconductor laser (Roithner RLT6505G),
emitting at a nominal wavelength of 650~nm with a threshold current of
20.1~mA.
Its temperature is set to $24.00\pm0.01\;^{\rm o}$C. The laser
output is collimated by an antireflection-coated laser-diode objective.
An external mirror is placed 60~cm away from the front facet of 
the solitary laser, which corresponds to a feedback time of 4~ns.
The threshold reduction due to the feedback is 9.4\%.
Throughout the paper, the injection current is set to 1.09 times the
solitary laser threshold.

\begin{figure}[htb]
\centerline{
\epsfig{file=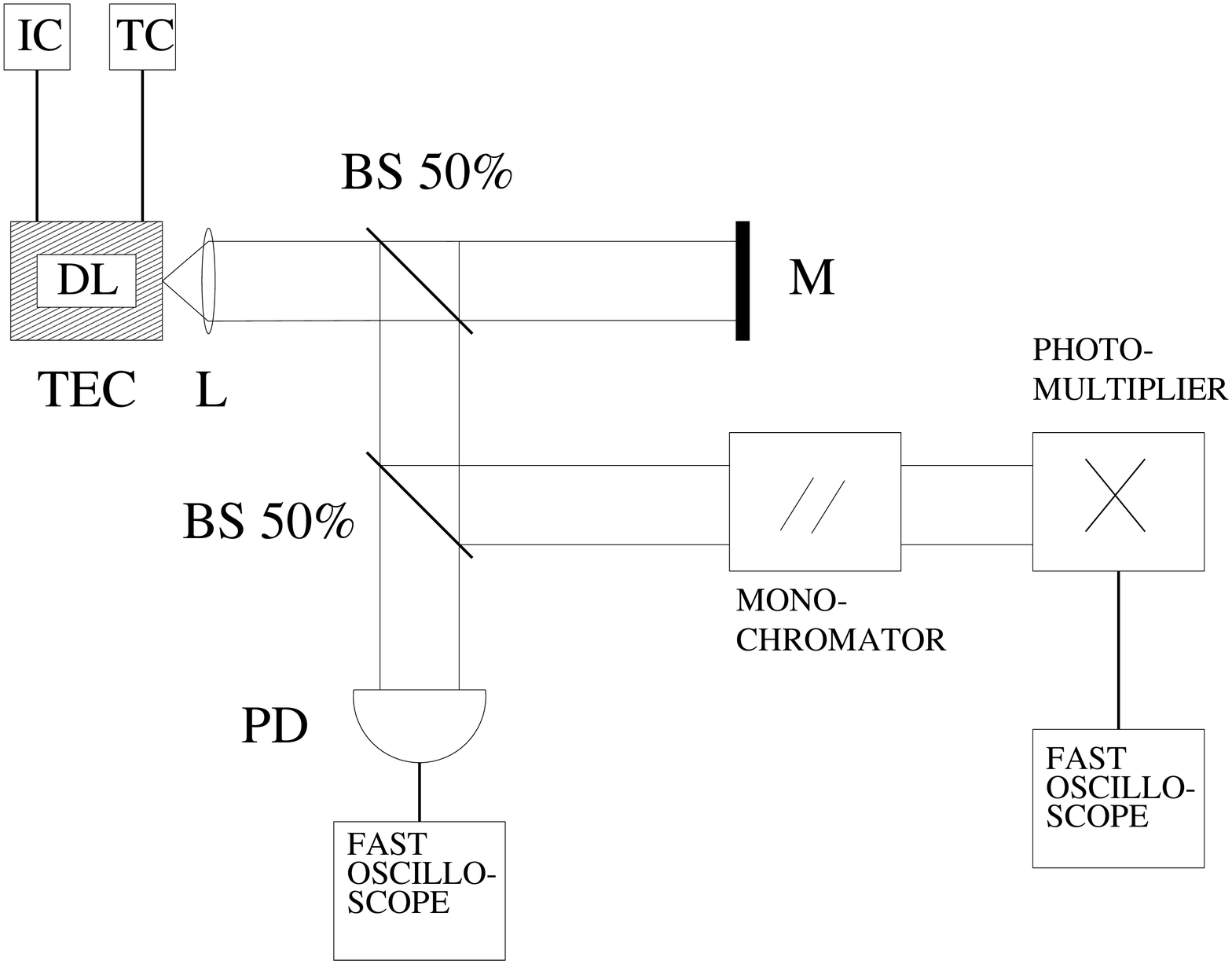,width=70mm}
}
\caption{
Experimental setup: LD, laser diode; BS, beam splitter; M, external mirror;
TEC, laser diode mount; PD, photodiode; IC, intensity controller;
TC, temperature controller.
}
\label{fig:setup}
\end{figure}

Part of the total output intensity is detected by a fast
photodiode and sent to a 500~MHz-bandwidth HP~54720D digital oscilloscope.
The rest passes through a $1/8m$ CVI monochromator
with a resolution better than 0.2~nm,
used to select the laser modes, and whose output is sent to a
Hamamatsu PS325 photomultiplier.
The photomultiplier signal is also recorded by the oscilloscope. 

The optical spectrum of our solitary laser shows at least ten active
longitudinal modes, with its maximum located at $\sim$658.4~nm and
a FWHM of $\sim$0.9~nm. When the feedback is turned on, the spectrum
broadens (up to a FWHM of $\sim$1.3~nm), and its maximum
becomes shifted $\sim$0.5~nm towards higher wavelengths
\cite{claudio,javier}.
For the feedback parameters chosen, mentioned above, the laser emits
in the low-frequency fluctuation regime. In this regime, we have
compared the dynamical behavior of the total emitted intensity of
the laser with that of the main mode of the laser with feedback (MMF),
and with a longitudinal side mode corresponding to the original main mode
of the solitary laser (MMS). The typical behavior is displayed in Fig.
\ref{fig:delay},
which compares the total intensity evolution with either that of the MMF
(traces a-b) or the MMS (traces c-d).
It can be seen that a power dropout is associated to an abrupt decay
of the former and a sudden activation of the latter. Note that the recovery
of the MMF is much slower than that of the total intensity, a fact that
has been already reported in the literature \cite{huyet,wallace}.
The activation is seen not to be symmetric, i.e. it does not occur
in the other side of the spectrum. We have found similar behavior
in other semiconductor lasers of similar quality, including
nearly-single-mode lasers.

\begin{figure}[htb]
\centerline{
\epsfig{file=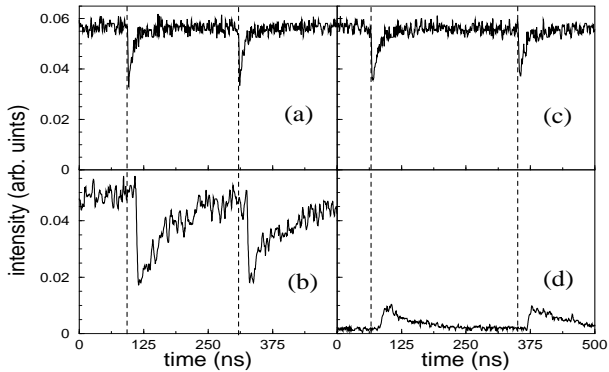,height=50mm,width=80mm}
}
\caption{
Modal structure of a dropout. Total intensity evolution (a,c) compared
to that of the main mode of the laser with feedback (b) and
of the original main mode of the solitary laser (d). Traces (a,b) and (c,d)
have been acquired simultaneously
(but note the intrinsic delay of the mode-selecting path of the setup -- see
text). Vertical dashed lines are a guide to the eye.
}
\label{fig:delay}
\end{figure}

We note that, even though the pairs of measurements (a,b) and (c,d) in
Fig. \ref{fig:delay} were acquired simultaneously, the time traces
exhibit a systematic delay of $\sim$20~ns between the
total-intensity dropouts and the corresponding modal powers (see vertical
dashed lines in the figure). This delay is spurious, due to the
electronic response time of the photomultiplier used in the mode-selecting
path of the experimental setup (Fig. \ref{fig:setup}), which is
substantially larger than that of the photodiode used to measure the
total intensity. However, as we will show in what follows, a closer
inspection of these results reveals
that this spurious delay is slightly {\em larger} for the MMS activation
than for the MMF dropout. Since both of these signals are measured
with the same detector, this observation leads to the conclusion that
the side-mode activation does not occur simultaneously to (nor before)
the dropout, but {\em after} it.

In order to estimate the delay between each dropout and its MMS activation
we proceed as follows. First, several (typically 40) time-trace pairs
containing a single
total-intensity dropout and its simultaneously measured modal event (either
MMF dropout or MMS activation) are averaged using a predefined event (a
given decay of the total intensity in our case)
as a trigger. In this way,
we average out fluctuations before the dropout event and during the
subsequent build-up, and refer all the time traces to a common time origin
(given by the predefined event mentioned above). The result of this
procedure is shown in Fig. \ref{fig:average}(a). One can already see in
this figure, which shows several averaged sets for each one of the
three quantities measured (total intensity, MMF intensity, and MMS
intensity), that the MMS activation occurs somewhat later
($\sim$1~ns) than the
MMF dropout (see vertical dashed lines in the figure). In order to
identify such a delay more clearly, we compare in Fig. \ref{fig:average}(b)
the MMS signal to the inverted MMF one. The delay becomes now evident.
Note also that the escape trajectories of the two modes (from the lasing
state in the MMF case, and from the off state in the MMS case) are 
basically parallel, which indicates that the instability mechanisms
are the same, and hence a direct comparison between them can be made.

We estimate the delay between the dropout and the side-mode activation
as the distance between the two corresponding parallel escaping trajectories,
which can be clearly identified in Fig. \ref{fig:average}(b) as two distinct
sets of straigth lines with the same positive slope. We perform a
piecewise local linear fit of each one of the averaged MMF and MMS time
series, and identify the time instants at which the slope takes its
maximum value. Figure \ref{fig:histogram} represents the distribution
of these times, for both the MMS activation and the MMF dropout, computed
from an statistics of 3000 dropout events. The distribution functions of
these two quantities are clearly separated, with a time difference
between their two mean values of 1.5$\pm$1.1~ns.

\begin{figure}[htb]
\centerline{
\epsfig{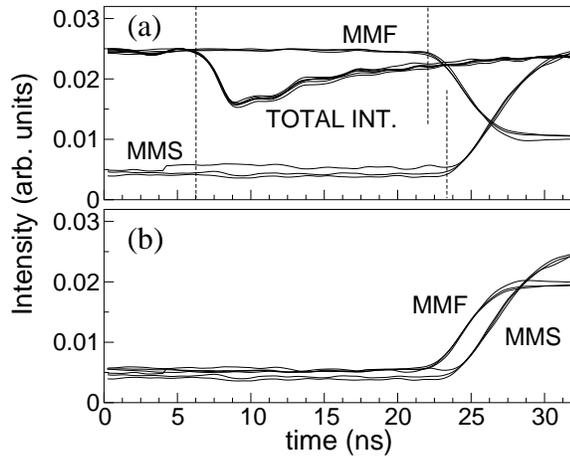}
}
\caption{
Averaged time traces of a dropout in the total intensity, main mode
of the laser with feedback, and main mode of the solitary laser.
In plot (b), the traces of the MMF have been inverted.
}
\label{fig:average}
\end{figure}

\begin{figure}[htb]
\centerline{
\epsfig{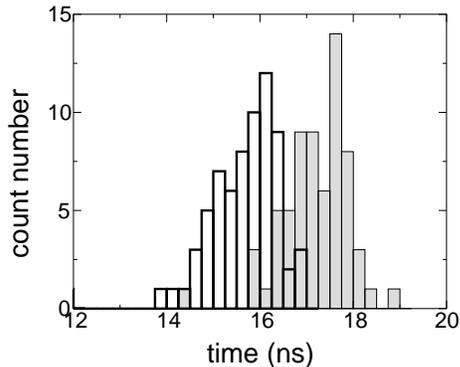}
}
\caption{
Distribution of times of occurrence of both the dropouts of the
main mode of the laser with feedback (white bars) and the activations
of the side mode -- main mode of the solitary laser (grey bars).
}
\label{fig:histogram}
\end{figure}

In conclusion, we have experimentally shown that low-frequency fluctuations
in a multimode semiconductor laser with global (i.e. non-frequency selective)
optical feedback are associated to sudden activations of a longitudinal
side mode
corresponding to the main mode of the solitary laser. These activations
are seen to occur after the dropouts of the main mode of the laser
with feedback, and hence after the total intensity dropouts of the
system. Therefore, in this case the side-mode activation cannot
account for the destabilization giving rise to the low-frequency
fluctuations. On the contrary, one can conjecture that the activations
are a natural consequence of the loss of power in the main modes of
the laser with feedback. Work directed to the theoretical modeling
of these phenomena is in progress \cite{javier}.

We thank J. Martorell and R. Vilaseca for lending us part of the
experimental setup.
We acknowledge financial support from Ministerio de Ciencia y
Tecnologia (Spain), under projects PB98-0935 and BFM2000-0264,
from the EU IST network OCCULT, and from the Generalitat de
Catalunya, under project 1999SGR-00147.

}
\end{multicols}

\end{document}